# Femtomolar detection of the heart failure biomarker NT-proBNP in artificial saliva using an immersible liquid-gated aptasensor with reduced graphene oxide

Stefan Jarić[a#*], Anastasiia Kudriavtseva[b#], Nikita Nekrasov[b#], Alexey V. Orlov[c], Ivan A. Komarov[b], Leonty A. Barsukov[b], Ivana Gadjanski[a], Petr I. Nikitin[c], and Ivan Bobrinetskiy[a,b*]

[a]BioSense Institute - Research and Development Institute for Information Technologies in Biosystems, University of Novi Sad, Novi Sad, 21000, Serbia

[b]National Research University of Electronic Technology, Moscow, Zelenograd, 124498, Russia, 8141147@gmail.com

[c]Prokhorov General Physics Institute of the Russian Academy of Sciences, 119991, Moscow, Russia; petr.nikitin@nsc.gpi.ru

**\*Corresponding authors: e-*mail*: sjaric@biosense.rs

# these authors contributed equally

**Abstract**

Measuring NT-proBNP biomarker is recommended for preliminary diagnostics of the heart failure. Recent studies suggest a possibility of early screening of biomarkers in saliva for non-invasive identification of cardiac diseases at the point-of-care. However, NT-proBNP concentrations in saliva can be thousand time lower than in blood plasma, going down to pg/mL level. To reach this level, we developed a label-free aptasensor based on a liquid-gated field effect transistor using a film of reduced graphene oxide monolayer (rGO-FET) with immobilized NT-proBNP specific aptamer. We found that, depending on ionic strength of tested solutions, there were different levels of correlation in responses of electrical parameters of the rGO-FET aptasensor, namely, the Dirac point shift and transconductance change. The correlation in response to NT-proBNP was high for 1.6 mM phosphate-buffered saline (PBS) and zero for 16 mM PBS in a wide range of analyte concentrations, varied from 1 fg/mL to 10 ng/mL. The effects of transconductance and Dirac point shift in PBS solutions of different concentrations are discussed. The biosensor exhibited a high sensitivity for both transconductance (2 µS/decade) and Dirac point shift (2.3 mV/decade) in diluted PBS with the linear range from 10 fg/mL to 1 pg/mL. The aptasensor performance has been also demonstrated in undiluted artificial saliva with the achieved limit of detection down to 41 fg/mL (~4.6 fM).

## 1. Introduction

Cardiovascular diseases (CVDs) are considered as the main cause of death worldwide (*World Health Organization. (2018)*). Therefore, fast detection of the CVD biomarkers is an urgent problem, on high demand especially in point-of-care (PoC) systems, where rapid diagnosis can make a significant difference in patient's outcomes. Among CVD markers, the N-terminal pro-B-type natriuretic peptide (NT-proBNP) has recently gained great interest as an early-stage marker of the heart failure (HF). NT-proBNP is released by the heart in response to increased pressure and volume overload, making it a reliable biomarker for assessing cardiac function. Its level in the blood correlates with the severity of the heart failure and can aid in identifying patients at higher risk or in need of immediate medical attention. One of the key reasons for its importance is its prolonged half-life of 60-120 min in blood, significantly longer than the half-life of BNP (B-type natriuretic peptide) that is 20 min in blood (Ben Halima et al., 2023).

Recently, the use of saliva was suggested for the control of NT-proBNP concentration as a noninvasive and less risky sampling screening approach (Foo et al., 2012). The concentration of NT-proBNP in saliva of HF patients with admission was found to be nearly a thousand time lower than that in blood, with median values of 5 pg/mL (2–10 pg/mL) and 3.5 ng/mL (1.5–10 ng/mL), respectively (Bellagambi et al., 2021). Hence, it is challenging to achieve the ultimate limit of detection (LOD) down to femtomole for relatively short assay time of this biomarker. Numerous label-free methods were suggested to address these problems. Basically, the cardiac-specific monoclonal antibody to NT-proBNP was used as a bioreceptor is such biosensors (Munief et al., 2019; Orlov et al., 2022). However, in general, antibodies used as bioreceptors suffer from parameter variation.

To the best of our knowledge, up to now only one high-affinity aptamer to NT-proBNP called N20a (Sinha et al., 2018) was suggested and investigated in the assembly of high-mobility transistors (Tai et al., 2019), and microfluidics sensors (Sinha et al., 2019). High affinity and fast response for NT-proBNP biomarker in the clinical range of concentrations were demonstrated. In spite of rapid response of such transistor-based sensors, LOD was slightly above the threshold value of 100-125 pg/mL in blood defined for healthy people (Munief et al., 2019). The N20a aptamer is of great interest to study the binding process on the subthreshold levels of HF markers because of high affinity to NT-proBNP and complex secondary structure, caused by its long backbone. Recently, we showed that graphene-based transistors can provide a powerful platform for investigation of interactions between different substances and specific aptamers attached to graphene surface (Nekrasov et al., 2022a). Our platform provides simple, accurate and fast measurements of HF biomarkers. Such performance, which is especially demanded as doctor's decision support systems during acute HF or surgery, is not achievable by the test systems based on electrochemical detection (Ruankham et al., 2023). A graphene-based field-effect transistor photochemically modified with aptamer was used for the fast detection of NT-proBNP with relatively high sensitivity (Nekrasov et al., 2022b). The use of graphene oxide (GO) for the enhancement of sensor performance is mainly attributed to the increase of immobilization density (Gao et al., 2022) due to higher roughness of the GO surface. Reduced graphene oxide–based field effect transistors (rGO-FETs) provide the same effect on sensitivity as graphene but with less efforts for the transfer process and preparation of conductive transistor channels, compared to the CMOS-based technology (Ben Halima et al., 2023). Recently, a simple graphene oxide based technology for production of GO-FET was suggested for biosensors development (Aspermair et al., 2021). In this paper, for the first time, we demonstrate the use of NT-proBNP specific aptamer (N20a) as a bioreceptor in a rGO-FET-based aptasensor that provides LOD down to 41 fg/mL (~4.6 fM) for NT-proBNP registration in artificial saliva.

## 2. Materials and methods

### 2.1 Materials and reagents

2 mg/mL graphene oxide suspension in water was purchased from Sigma Aldrich (USA), catalog number 763705. GO in solution is presented in the form of a monolayer sheet with size less than 10 µm. N-methyl pyrrolidone (NMP) was obtained from Sigma Aldrich (USA). Frozen solutions of 1.18 mg/mL of NT-proBNP in 10 mM potassium phosphate, 150 mM NaCl, pH 7.4, and 0.7 mg/mL of cardiac troponin I (cTnI) in 10 mM HCl were purchased from HyTest (Moscow, Russia). Molecular mass of NT-proBNP of 8589 Da was confirmed by MALDI-MS. N20a aptamer with a sequence 5' – GGC AGG AAG ACA AAC AGG TCG TAG TGG AAA CTG TCC ACC GTA GAC CGG TTA TCT AGT GGT CTG TGG TGC TGT - 3' (Sinha et al., 2018) with amino-modified 5' end, purified using denaturing polyacrylamide gel electrophoresis was purchased from DNA Synthesis, LLC (Moscow, Russia). 1-Pyrenebutyric acid N-hydroxysuccinimide ester (PBASE) was acquired from Sigma Aldrich (USA). Ethanolamine (ETA) was acquired from Sigma Aldrich (USA). Phosphate-buffered saline (PBS, 10X solution) was purchased from Fisher Bioreagents (USA). Bovine serum albumin, lyophilized powder (BSA) was obtained from Sigma Aldrich (USA). Dimethylformamide (DMF) and isopropyl alcohol (IPA) were purchased from Component-Reactive (Russia) and Sigma Aldrich (USA).

## 2.2 Fabrication of the rGO-FETs chips

Two types of rGO-FETs were investigated: GO monolayer deposited on commercially available Interdigitated Electrodes (IDE) and arrays of lithographically patterned GO stack channels.

For GO monolayer deposition on predefined electrodes we used two types of IDEs, G-IDE222 (Drop Sens, Spain) and Micrux ED-IDE1-Au (Micrux, Spain), and varied the deposition parameters of GO film (see Supporting Information, Fig. S1-S3). The working device was produced by suspending graphene oxide in water/NMP mixture (10/90 %) to get the concentration of 0.2 mg/mL (pH=7.3) and drop casting on interdigitated area. Prior to the deposition, the glass surface between electrodes was activated by APTES (2% v/v in ethanol) in 1 h incubation followed by 2 h thermal annealing at 120 °C on a hotplate, while the gold surface was modified with cysteamine (0.1 M in water) in 2 h incubation. The size of a single GO flake is less than the distance between the gold electrodes. Thus, glass modification is needed to develop a GO-based channel of field-effect transistor. Finally, 40 µL of GO suspension was drop-casted over the IDE for 2 h incubation, followed by a 30-min annealing at 100 °C to evaporate most of the water/NMP solvent residuals. GO was reduced using hydrazine vapor at 80 °C for 2 h and then additionally thermally annealed at mild parameters (200 °C for 1 h) to avoid damaging the gold electrodes. Arrays of rGO-FETs on the 4-inch silicon wafers were provided by Dr. Kireev (The University of Texas at Austin, USA), and details on the preparation are given in the Supporting Information.

## 2.3 Assembling of the rGO-FET aptasensor

The G-IDE222 electrodes were chosen for the aptasensor assembly due to lower leakage current demonstrated in our measuring set-up (Fig. S3). The rGO-FET-based aptasensors were assembled in a similar way as it was previously described for graphene devices (Nekrasov et al., 2022a). PBASE linker was immobilized on graphene oxide surface by 3 h incubation of a 5 mM PBASE in DMF under -0.3V potential applied to an auxiliary electrode (AE) (Fig. 1) to increase the density of the linker (Hao et al., 2020). The rGO-FET was then rinsed consequently with DMF, IPA, and deionized (DI) water to remove reagent excess. A 100 µl drop of 100 nM of N20a aptamer solution in 1xPBS (with pH = 7.4) was introduced into the well mounted on a rGO-FET chip and kept overnight in humid atmosphere to ensure aptamer binding to the PBASE linker. After rinsing several times in PBS solution to remove non-bound aptamers, 100 mM ethanolamine solution in PBS was introduced and kept for an hour in the well to block and deactivate non-bonded reactive groups. To block remaining rGO surface a 0.5% BSA aqueous solution was incubated on rGO-FET IDE chip for 30 min.

## 2.4 Aptasensor characterization and NT-proBNP detection

Atomic force microscopy (Solver-PRO, NT-MDT, Russia) was used to estimate the GO thickness and surface roughness. Profilometry of the IDE was performed on a DektakXT stylus profilometer (Bruker, USA). The graphene oxide surface was investigated by microRaman spectroscopy (Centaur HR, Nanoscan Technology, Russia) with a 100x objective (NA=0.9) at a 532-nm wavelength (Cobolt, Solna, Sweden) with a beam spot of 1 µm$^2$ and laser power of 0.5 mW. For liquid gate measurements, we introduced the Ag/AgCl pellet electrode (Science Products GmbH) with a diameter of 1 mm in a droplet of the analyte solution. Electrical measurements were performed on HP 4145A (Yokogawa-HP, Japan) and IPPP 1/5 (MNIPI, Belarus) semiconductor parameter analyzers. Gate current was also monitored on the level of leakage current from the GO channel to the gate electrode.

PBS of 160 mM (1x concentrated) was diluted to 0.1x and 0.01x PBS, which reduced ionic strength to 16 mM and 1.6 mM, respectively. Intermediate concentrations were prepared by respective mixing of buffer salts (see Supporting Information for details, Fig. S5). Each measurement with different ionic strength was performed with 15 µl of buffer inserted with a micropipette. After each set of measurements with one buffer (ionic strength value), the washing process was done with DI water, precisely, three times of 45 µl of DI water in PDMS well followed by 10 min of chip drying at room temperature (20-25 °C).

Several solutions of different concentrations in the range of 1 fg/mL to 10 ng/mL of NT-proBNP were prepared in 0.1x PBS and 0.01x PBS. For $I_d$-$V_g$ measurements, we first put 80 µL of pure 0.1x PBS in a PDMS well on rGO-FET. Then we replaced it with the NT-proBNP spiked solution and incubated for 10 min. The measurements were repeated for all concentrations of HF biomarker. Artificial saliva (AS) was ordered from "Apoteka Beograd" (Belgrade, Serbia). AS is composed of carboxymethyl cellulose and produced according to the recipe, registered under the Republic of Serbia's master preparations. NT-proBNP was dissolved in AS the same way as described above for PBS.

## 3. Results and discussion

### 3.1. Characterization of rGO-FETs

Two different methods based on the spin-coating and drop-casting were used for deposition of the monolayer GO film on IDEs. The structure produced by the drop-casting has demonstrated higher operating current and symmetric transfer characteristics (Fig. S1b). The optical image of the IDE chip with the rGO channel on working electrodes (WE) is presented in Fig. 1A. We measured current voltage characteristics (CVCs) for liquid gate configurations of the IDE chips with different gold electrode thicknesses: 230±10 nm (ED-IDE1-Au) and 120±10 nm (G-IDE222). Both types of devices demonstrate good leakage characteristics that enable using it as immersible sensor. The rGO-FET based on the G-IDE222 electrodes demonstrated lower leakage current and higher threshold voltage (Fig. 1B and Supporting Information, Fig. S3), so it was used for the aptasensor assembly. After hydrazine reduction we didn't observe significant improvement of channel conductivity, so additional thermal reduction was used. The output electrical properties of all tested devices were linear (Supporting Information, Fig. S1) with resistance less than 1 kOhm for rGO-FET. The Dirac point had positive values, indicating p-doping of the reduced graphene oxide channel. The FET channel consists of a monolayer of rGO flakes with an average film thickness of 1.7±0.3 nm (Fig. 1C, and Supporting Information, Fig. S2). The estimation of thickness is limited by the high roughness of the supporting glass substrate (1.3±0.3 nm) and gold electrodes (3.8±0.3 nm). A possible competitive effect between APTES and cysteamine can hinder the linkage of GO to the IDE surface. We assume that the film consists mainly of interconnected individual monolayer flakes with lateral sizes of several µm, each forming continuous channels in 10-µm-wide gap between gold electrodes. The Raman mapping of G band intensity (Fig. 1D) confirms the uniformity of rGO film observed on AFM with small areas without rGO and small islands of multiple layers of rGO. Notably, the intensity of Raman bands of rGO increased on gold electrodes due to plasmonic enhancement.

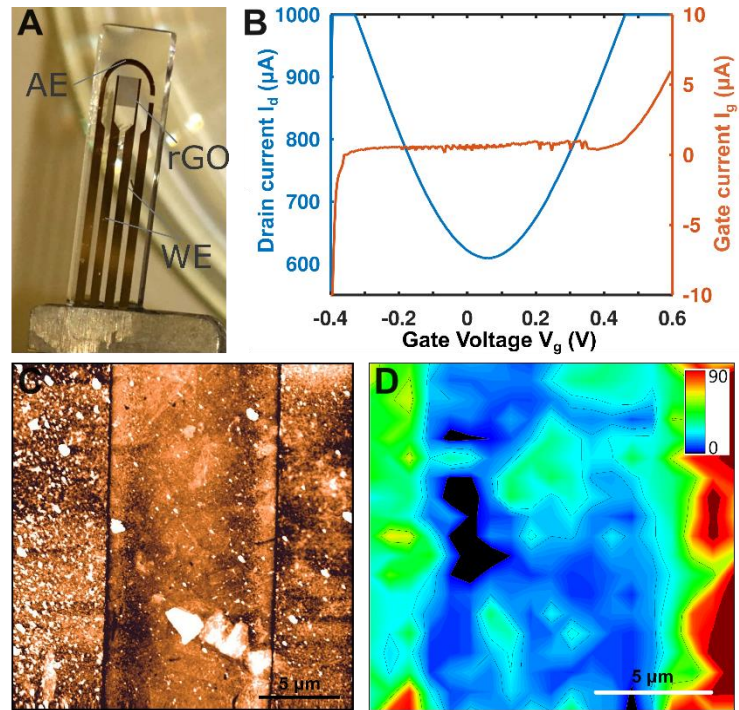

**Fig. 1.** Characterization of rGO-FET formed on G-IDE222 electrodes. (**A**) Photo of rGO-FET aptasensor on an immersible IDE chip. (**B**) Transfer characteristic (blue) and leakage (gate) current (orange) of rGO-FETs on IDEs in the liquid gated configuration ($V_{ds}$=0.1 V). (**C**) Atomic force microscopy image of a rGO channel. (**D**) Raman mapping of G-band intensity of rGO film across electrodes. Scale bar for (C) and (D) is 5 μm.

*3.2 Effect of Ionic strength on rGO-FET properties*

The ionic strength of a solution is critical for the biosensor's performance. Its main effect is in the Debye length which limits the sensitivity of FET-based sensors to the large sizes of analyte molecules or receptors (Kesler et al., 2020). To investigate the effect of ionic strength in the liquid gate measurements, we used an array of rGO-FETs with 20×20 μm² channel in undiluted buffer, where one rGO transistor on the same chip was set as a gate electrode. The rGO-FETs characteristics in the array were linear as for the IDE structure with relatively good reproducibility (Supporting Information, Fig. S4). That provided the symmetric structure of the capacity formed by the electrical double layer (EDL). The EDL structure consists of a first layer (Stern layer), which is formed by ions of the electrolyte adsorbed on the channel's surface, and second layer formed by opposite charge ions attracted to the surface charge via Coulomb force (outer Helmholtz layer), electrically screening the first layer (Kesler et al., 2020). The thickness of the EDL is Debye length, which is reciprocally proportional to the square root of the ion concentration (for details, see Supporting Information and Fig. S5). For determination of the Debye length, the relative permittivity of 1x PBS (ionic strength - 160 mM) was used as 80. For the 1x PBS, the Debye length was ~0.79 nm.

We performed liquid-gated measurements of $I_d$-$V_g$ by varying ionic strength of PBS (Fig. 2A, S6a). The potential difference between the interface of the bulk solution and the solid carbon modulates the carrier density in the rGO layer in both electron and hole transport regime (Munief et al., 2019). The Dirac point showed steady dependence on the ionic strength increase (Fig. 2B) typical for graphene-based FETs (Rodrigues et al., 2022). For bare rGO-FETs, the log-linear left shift of the Dirac point with ionic strength increase is different from that observed previously for rGO-FET (Munief et al., 2019) and correlates with bare graphene FETs (Ping et al., 2017). Such behavior can reflect proper reduction of GO by combining oxygen species removal with low defect density in rGO providing higher sensitivity of charge carriers to ionic screening in the Stern layer. The left shift of Dirac point (n-doping) with increasing of ionic strength corresponds to the increase of ionic screening of the charges formed by negative ions in the vicinity of graphene surface (Stern layer) (Hwang et al., 2020; Ping et al., 2017). Interestingly, the absolute current at the electron-doping branch grows faster than current at the hole-doping branch (Fig. S6b) due to the additional electron doping of rGO with an increase of ion concentration in the Stern layer. Similar behavior was observed in the transconductance growth with increasing the ionic strength related to the electrostatic effect increase (for details, see Supporting Information, Fig. S6c). A variation in ionic strength of solution affects the electrical properties of liquid-gated FETs as gating is managed through the buffer ions. Low ion concentrations decrease transconductance in graphene-oxide sheets (Rodrigues et al., 2022). The same mechanism responsible for changes of both Dirac point and transconductance is reflected by the Pearson correlation coefficient (PCC) almost close to -1 (Fig. S6d).

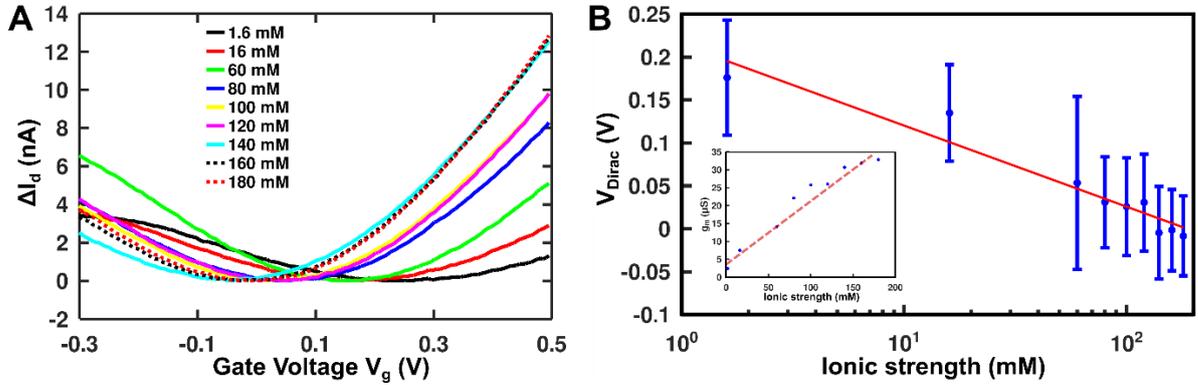

Fig. 2. The $I_d$-$V_g$ curves with normalized current for different values of PBS ionic strength (**A**) and shift of the Dirac point ($V_{ds}$=40 mV), N=2 (**B**) (Inset: dependence of transconductance (at $V_g$ = 0.3 V) on ionic strength). The error bars are calculated as a standard deviation of the mean value of the Dirac points.

The rGO film has a more irregular and rougher surface (Fig. S4b) that is of the order of the EDL thickness in 0.1x PBS. Such rough surface can increase the Debye length (Hwang et al., 2020; Zheng et al., 2021). The sample-to-sample variations of Dirac point (Fig. 2B) were high because of the relatively thin film with high roughness that could alter the model of a perfect flat capacitor used in our estimation.

*3.3 Dirac point and $g_m$ variations during rGO-FET aptasensor assembly*

The rGO-FET chips were assembled into the aptasensor as described in *Materials and methods* and were used to analyze electrical response to NT-proBNP in PBS with different ionic strengths. The assembling of rGO-based aptasensor was performed in a way similar to previously published for graphene-based FETs (Nekrasov et al., 2019, 2022a). We observed some differences in characteristics of developed sensors mainly because of rGO film structure and different substrate. PBASE attachment remarkably affects the electrical properties of rGO-FET resulting in both transconductance increase and Dirac point shift (Fig. 3A). Similar but less pronounced effect we have observed also for rGO-FETs array with stack of GO layer (Supporting Information: Fig. S7). The positive shift in Dirac point is typical for PBASE absorption on graphene by π-π stacking and associated with p-doping from NHS group (Wu et al., 2017). The transconductance increase for both branches was also observed for graphene modified with pyrene-based surface ligands (Mishyn et al., 2022), and we associate such behavior with ionic redistribution in the Stern layer that can modulate the thickness of EDL by planarization of rGO surface. Increase of capacitance of rGO due to a coupling with PBASE aromatic rings as predicted for stacked layers (Cui et al., 2021) can also give inputs to the transconductance growth. The transconductance $g_m$ in linear curve of liquid-gated graphene based FETs can be calculated using semi-classical Drude model (Vasilijević et al., 2021) :

$$g_m(V_{ds}, V_{Gg}) = I_d(V_{ds}, V_g) \cdot \frac{C_{tot}}{e \cdot n}, \qquad (1)$$

where $I_d$ is the drain current at linear part of CVC, $C_{tot}$ is the total capacity of rGO-FET channel (see Supplementary Information for calculations) and $n$ is the charge carrier density per cm². The Dirac point shift $\Delta V_{Dirac}$ was calculated as $\Delta V_{Dirac}$= $V_{Dirac}$-$V_{Dirac/rGO}$, where $V_{Dirac}$ – is the measured Dirac point and $V_{Dirac/rGO}$ – is the Dirac point of the bare rGO-FET. Transconductance change $\Delta g_m$ was calculated as $\Delta g_m$= $g_m$-$g_{m/rGO}$, where $g_{m/rGO}$ is the transconductance of the bare rGO-FET.

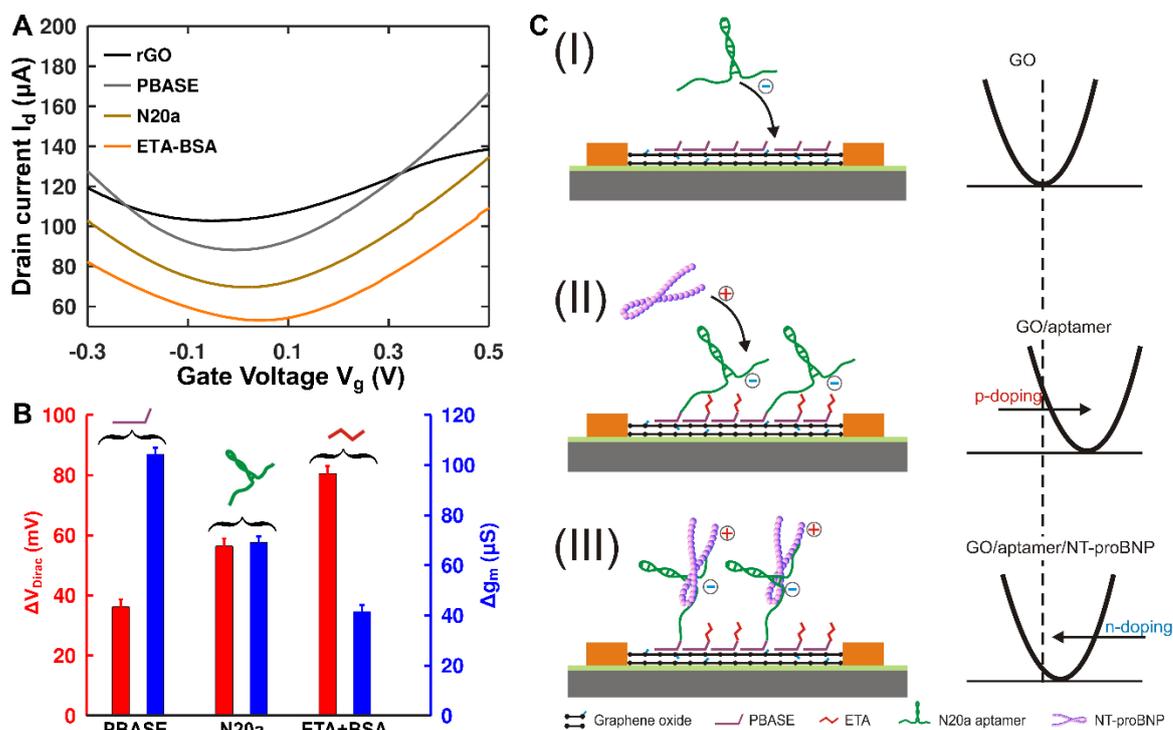

**Fig. 3.** NT-proBNP biosensor assembly. (A) $I_d$-$V_g$ curves measured in 1x PBS with an Ag/AgCl electrode after each step of assembly ($V_{ds}$ = 0.1 V). (B) Dirac point shift and transconductance change (at $V_g$ = -0.2 V) for each step of sensor assembly. (C) Schematics of rGO-FET electrostatic doping by aptamer covalent binding (I-II) and aptamer reconfiguration in presence of NT-proBNP (II-III).

Aptamer binding to rGO/PBASE surface resulted in the decrease of conductivity and Dirac point positive shift to 56±5 mV (fig. 3B). This is quite consistent with the model of aptamer negative charges bringing close to graphene surface causing electrostatic p-doping of the graphene plane. Nevertheless, different papers provide different results of aptamer binding due to the competitive effect of electrostatic doping and direct charge transfer between aptamer and graphene surface (Danielson et al., 2020; Nekrasov et al., 2022a). A strong change in conductivity response is observed due to the larger number of nucleotides (72 bases), resulting in a large accumulative negative charges from the phosphate groups that are brought close to the graphene surface (Fig. 3C). The negative-charged backbone results in overall p-doping of rGO. One may conclude that direct charge transfer between aptamer and rGO is less favorable due to the specific configuration of the sensor surface. Interesting to note, while the total amount of charge carriers has increased, the transconductance for both hole and electron branches has decreased. This can be related to the decrease in mobility of the main charge carriers because of molecular absorption. It should be noted, that even reduced GO still has better hydrophilic properties compared with graphene that can result both in water sorption during the storage and less favorable hydrophobic binding of biomolecules. After each step of aptasensor assembly with biomolecules (N20a, BSA) we observed p-doping with a positive Dirac point shift, as well as a transconductance decrease . While p-doping effect should increase the conductivity of the graphene oxide channel, the increase of charge carrier scattering due to adsorbed molecules can greatly decrease the charge carriers mobility in the channel (Ciou et al., 2023).

3.4. *Sensing performance toward NT-proBNP in buffers with different ionic strength*

To control the sensing performance of rGO-FETs, we prepared a range of concentrations of NT-proBNP from 1 fg/mL to 10 ng/mL diluted in PBS with ionic strength of 1.6 and 16 mM. We have estimated whether 10 min incubation is enough for the sensor performance validation. For this purpose, the diffusion length for peptide in solution was calculated using the 3D random walk models:

$$L_D = \sqrt{6Dt}, \quad (2)$$

where $L_D$ - diffusion length, $D$ – diffusion coefficient, $t$ – walk time. We assume $D = 2*10^{-10}$ m$^2$/s (Torres et al., 2012) and the estimated diffusion length is L = 0.85 mm, which corresponds to ~10 μL volume of sample in the PDMS chamber used in this work. It means that in a distance of about 850 μm from the graphene surface we can assume all the peptide can achieve the rGO surface and can be conjugated with aptamer. Then the estimated maximum density of peptide adsorption is $2.5×10^4$ molecules/μm$^2$, which can roughly define the upper limit about 1-10 pg/mL of linear detection range.

When binding the NT-proBNP, we observed the left shift of Dirac point (Figure 4A,B) for both PBS dilutions. This agrees with the model of aptamer reconfiguration and decreasing the effect of charges on the graphene. Note, the initial position of Dirac point for assembled aptasensor shifts from 81±5 mV for 1x PBS (Fig. 3C) to 92±5 mV and 174±5 mV for 0.1x PBS and 0.01x PBS, respectively. This shift agrees with the observation of ionic strength effect on rGO-FET (Fig. 2) and confirms that the effect is preserved for rGO modified with PBASE and other biological molecules (Ping et al., 2017; Rodrigues et al., 2022). When the aptamer binds the peptide in the solution, its electrostatic effect on graphene diminishes resulting in a decrease of charge carriers' concentration. The Dirac point shifts to more negative values because of positive charges brought by the peptide (isoelectric point pI=8.5 for NT-proBNP in pH=7.4 of standard buffer solution (Chu et al., 2017)). The N20a aptamer (with mass ~22.5 kDa) is three times larger than NT-proBNP and the aptamer reconfiguration may strongly effect on doping. This effect is more pronounced for higher dilutions of the buffer, while in 0.1x PBS we observe a great variation in the Dirac point resulting in a high calculation error (Fig. 4C). The transconductance change provides a steadier dependence (Fig. 4D). The biosensor demonstrates high sensitivity for both transconductance (2 μS/decade) and Dirac point shift (2.3 mV/decade) in 0.01x PBS. Due to interaction between NT-proBNP molecules, Langmuir-Freundlich isotherms was used to describe the Dirac point shift and change in transconductance including molecule-molecule interaction parameter $x$=0.1 (Ayawei et al., 2017; Brazesh et al., 2021):

$$\Delta S = S_0 + A_0 \frac{KC^x}{1+KC^x}, \qquad (3)$$

where $\Delta S$ -is a signal change ($S = \Delta V_g$ or $\Delta g_m$), the constant $S_0$ represents non-specific adsorption (VanDer Kamp et al., 2005), $A_0$ – adsorption capacity constant, $K$ – affinity constant, $C$ – analyte concentration, $x$ – analyte-receptor interaction parameter ($x \approx 0$ for strong interaction, $x$=1 for weak interaction) (Koble and Corrigan, 1952). Therefore, molecules in solution are strongly interacting with each other, and that affects the binding efficiency with aptamer. Moreover, the binding kinetics between aptamer and analyte is strongly dependent on ionic strength and can overcome Debye screening length for undiluted buffer (Wang et al., 2019).

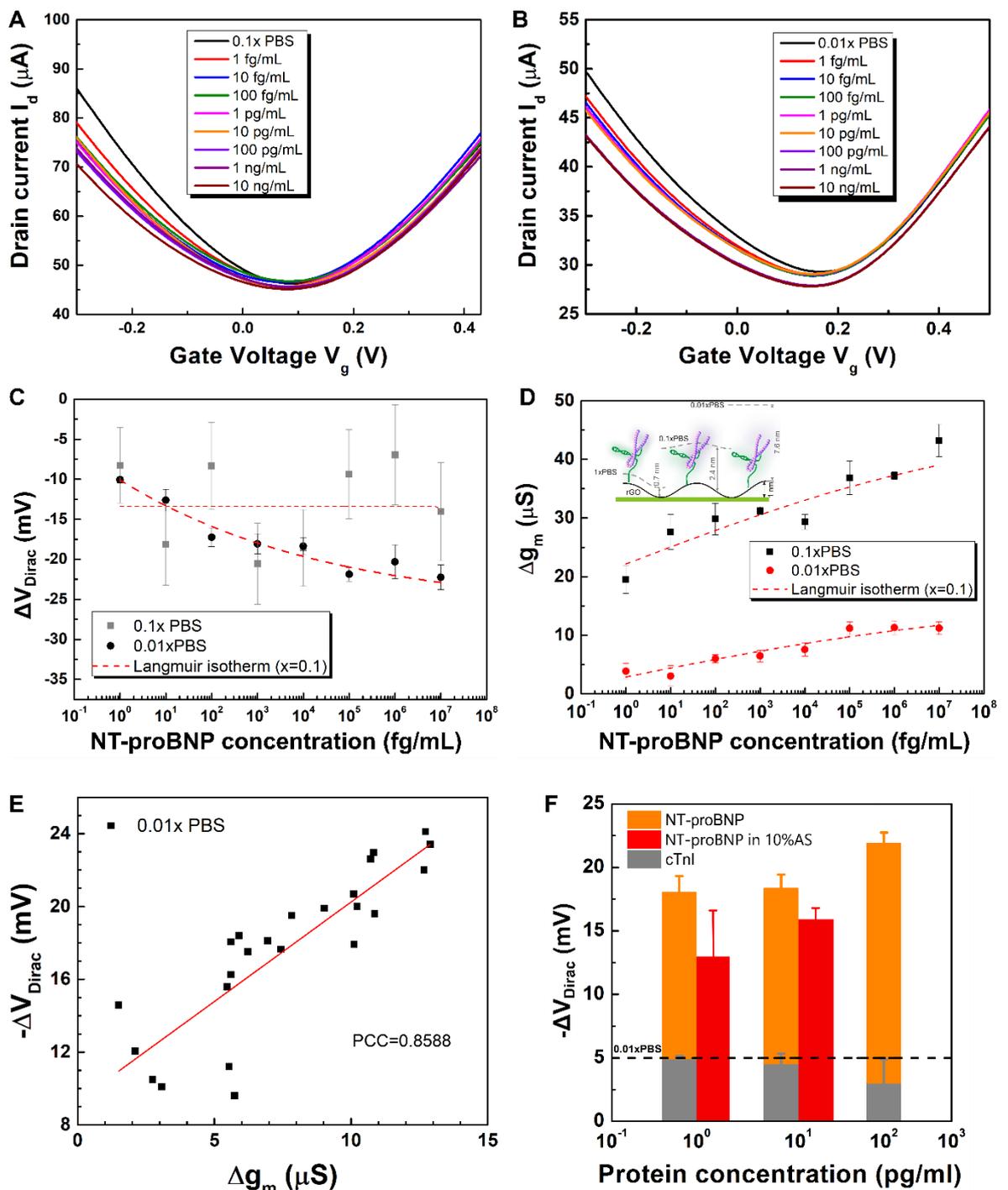

**Fig. 4**. $I_d$-$V_g$ curves of liquid-gated rGO-FET for different concentrations of NT-proBNP in (A) 0.1x PBS and (B) 0.01x PBS, $V_{ds}$=100 mV. (C) Change in Dirac point for increasing concentrations of NT-proBNP in PBS of different ionic strengths. Dashed lines are Langmuir isotherms. (D) Transconductance change with respect to NT-proBNP concentration variation, as extracted from the most linear part of hole branches of transfer curves at $V_g$ = -0.2 V (for 0.1x PBS) and $V_g$ = -0.15 V (0.01x PBS) (Inset: schematics of NT-proBNP binding to aptamer, dashed lines depict Debye length at different ionic strength). (E) Two-parameter rGO-FET response profiles: correlation between transconductance change and Dirac point shift during the NT-proBNP detection in 0.01x PBS. (F) The specificity test for NT-proBNP in 0.01x PBS and in 10% AS diluted in 0.01x PBS, in comparison with cTnI in 0.01x PBS. The black dash line represents the noise level, which was the Dirac point shift variation obtained by multiple measurements using the biosensor completely assembled and passivated with BSA in PBS for transfer characteristic curves.

By variation of ionic surroundings of rGO surface we can change the sign of the biosensors electrical response (Kim et al., 2013). Thus, we can have a competing process of negative charges from aptamers and positive charges from peptides. When discussing the interaction of peptides with graphene aptasensor, we have to take into account internal charges of the electrostatic field, which depends on the ionic strengths of the solution (Ping et al., 2017). Due to the presence of positively charged arginine and histidine amino acids (Hall, 2004) NT-proBNP can bring a net positive charge close to graphene surface. Thus, for higher ionic strength solutions, the charges on the peptide can affect the charge carriers in graphene only at a close vicinity to the surface (Fig. 4D, inset).

High roughness of the rGO channel can effectively increase the EDL thickness and channel sensitivity (Ganguli et al., 2020; Hwang et al., 2020). As far as aptamer is larger than the EDL layer thickness in 1x PBS, only a small portion of its backbone can affect the channel conductivity. Thus, a conformational reconfiguration of the aptamer can reflect the NT-proBNP concentration change similar to the model suggested for small molecules detection (Nekrasov et al., 2022a). On the other hand, the large size of the aptamer can cause higher signal variations. We suggest that an increase of EDL thickness may enhance the reliability of the sensing data.

We assume that the capacity varied due to the conjugation of the aptamer and NT-proBNP molecule, altering outer Helmholtz layer of the EDL (Figueroa-Miranda et al., 2022) and resulting in undetermined shift of Dirac point (charge neutrality point) in 0.1x PBS. Dirac point position mainly depends on double layer capacity and is directly associated with doping level. On the contrary, transconductance mostly depends on mobility of the charge carriers, total capacity, and doping level (See Supporting Information, Eq. S2). Since the scattering of charge carriers increases due to additional molecules' sorption, which results in charge carriers' mobility decrease, the transconductance decreases with NT-proBNP concentration, but with less pronounced change because of the smaller total capacity of diluted electrolytes. Therefore, a transconductance change is introduced as a responsive parameter, and a logarithmic increase is observed with the rise of analyte concentration (see Fig. 4D). These results are in agreement with a rGO-based biosensor that uses the current change as responsive parameter for NT-proBNP detection (Munief et al., 2019).

When decreasing ionic strength to 1.6 mM (0.01x PBS) the change of capacity caused by peptide binding is neglectable, and direct doping is observed by a clear Dirac shift with increasing NT-proBNP concentration. Figure 4E shows the two-parameter $\Delta V_{Dirac}$-$\Delta g_m$ response profiles for all NT-proBNP concentrations (for each set of three measurements) in solution of PBS with 1.6 mM ionic strength. The high linear pattern with PCC≈0.86 on this two-dimensional voltage-conductivity graph is clearly visible. For 0.1x PBS the PCC is close to zero meaning that there is no correlation between Dirac point shift and transconductance change (Fig. S8). The response in transconductance for 0.01xPBS is weaker in comparison with 0.1x PBS, which can be explained by a decrease in total capacity in GO-GFET (see Supplementary Information, Table S2). Recently, direct Hall effect measurements in biosensing experiments demonstrated the growth of charge carries concentration and a decrease of mobility with analyte concentrations (Ciou et al., 2023). Thus, when inputs from these two effects are similar in weight, the mutual compensation can lead to weak response of the biosensor. The PBS solution with 1.6 mM ensures that charged conjugates of NT-proBNP and aptamer are located within the Debye screening length of the PBS solution, estimated to 7.6 nm.

Control experiments were conducted for another main HF marker – cardiac troponin-I (cTnI). Three concentrations of cTnI were used for specificity test in 0.01x PBS (Fig. 4F) and the response was observed on the noise level of the buffer, indicating high specificity of the biosensor (Gao et al., 2022).
In the research, we used the same sensor multiple times to perform measurements in different solutions (0.1xPBS, 0.01xPBS, AS). To regenerate the rGO-FET biosensor between experiments, we used 10 mM NaOH due to its nucleic acid denaturating properties (Ruankham et al., 2023). Deformation of the aptamer chain mitigated by NaOH releases the NT-proBNP from the surface. That demonstrates that the sensor can be reusable.

## 3.5. Sensor performance in artificial saliva

Finally, we have demonstrated the performance of developed sensors for analysis of AS spiked with NT-proBNP. AS mimics the chemical composition and viscosity of real saliva (Qureshi et al., 2022). The ionic strength of saliva is lower than of blood and is considered to be several tenths of mM (Gal, 2001; Ruankham et al., 2023). We suggest a simple method that does not need any preparation of AS for measurements i.e. raw samples could likely be used in the real-life setting. The Dirac point of rGO in AS is -30 mV (Fig. S9a). Using the ionic strength calibration curve in Fig. 2B we can estimate that the AS ionic strength is higher than 60 mM, which agrees with the data for AS (Teubl et al., 2018). The total current decrease reflects a change of composition from PBS to AS, and impurities reduce the channel capacity (Xia et al., 2009). The best agreement with the calibration curve for NT-proBNP was demonstrated for concentrations in the range from 10 fg/mL to 1 pg/mL based on Dirac point shift (Fig. 5). We assume, there is no direct charge transfer from peptide to rGO channel at high ionic strength (fig. 4D, inset). In AS, EDL is very thin, and the bonding with peptide causes aptamer folding, which changes the charge concentration in the EDL. We observed the weak degradation of GFETs characteristics after several measured steps caused by the more viscous nature of AS. Notably, the higher variation in the sensor response was observed for increasing concentrations of NT-proBNP. We assume that this effect occurs due to the complex interaction between graphene, aptamer, and peptide.

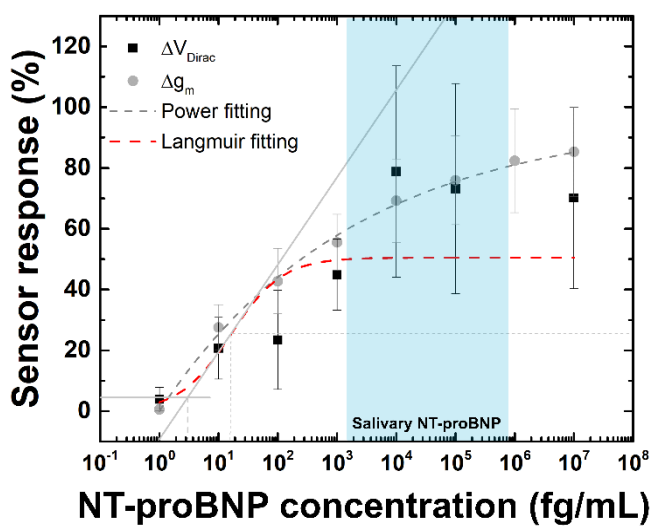

**Fig. 5.** Plot of the binding isotherm with different NT-proBNP concentrations in AS calculated for Dirac point shift (black squares) and transconductance change (grey circles). 100% corresponds to the maximum change in the signal at the highest concentration. The sensor covers the NT-proBNP salivary range of 5 to 700 pg/mL for HF patients (blue) according to (Bellagambi et al., 2021 and Foo et al., 2012). N=2

The results indicate that the lowest concentration of NT-proBNPs that can be detected in AS is in the range of 10–100 fg/mL. Above 10 pg/mL the saturation point was observed (Munief et al., 2019). The isotherm that describes the sensor response has a molecule-molecule interaction parameter $x$=1. This signifies that due to the higher ionic strength, NT-proBNP molecules have lower interaction in AS.

Using the adsorption isotherm curve, we estimated the LOD value with 41 fg/mL of NT-proBNP (Fig. S9b). These values are quite below the threshold NT-proBNP salivary range of HP patients (Bellagambi et al., 2021; Foo et al., 2012) and thus the sensor can be used for more detailed analysis of saliva from the patients with different HF stages as well as for HF-predictive screening procedures. Notably, using the diluted spiked AS samples the results are correlated with calibration data (Fig. 4F) reflecting decrease of salivary matrix effect on sensor performance.

In spite of outstanding limit of detection, obtained for NT-proBNP, the developed sensor provides higher dynamic range compared to those observed by other groups (Rodrigues et al., 2022). The range fits well the clinical range for NT-proBNP (Munief et al., 2019). The interplay of aptamer length when binding the peptide and the peptide size must be considered when choosing the analyte solution ionic strength. Using correlation analysis of two weakly dependent parameters like Dirac point shift and transconductance change, the reliability of a signal and performance of FET-based biosensor can be increased.

The proper consideration of correlation between salivary and blood NT-proBNP concentration, as well as differentiation from other possible diseases (like periodontal) is still needed in medical environment. The suggested approach for rGO-FET sensor development is scalable and can be used to produce low-cost disposable chips. Nevertheless, the suggested method needs more experimental validation from clinical application before transferring toPoint-of-Care technologies in clinical environment.

## 4. Conclusions

A novel aptasensor based on reduced graphene oxide FET for NT-proBNP detection has been demonstrated. The effect of different ionic strengths of the solution in which the sensor is placed on the sensor performance has been investigated. We have suggested two-dimensional voltage-conductivity correlation analysis to validate the performance of the aptasensor, where both receptor and analyte complexes can have a competitive effect in the electrical performance of rGO channel. For high PCC, the results of the aptasensor are more reliable and demonstrate high sensitivity. The Dirac point of the biosensor left-shifts linearly upon increasing the logarithmic concentration of the target NT-proBNP peptide (from 10 fg/mL to 100 pg/mL) for 0.01x PBS and correlates with transconductance decrease with coefficient 0.86. There is no correlation between transconductance and Dirac point for 0.1x PBS. The high specificity of aptasensors was demonstrated by measurements using another cardiac marker cTnI, where the aptasensor did not produce signal. The described device demonstrates unprecedented sensitivity to NT-proBNP with LOD as low as 41 fg/mL in artificial saliva. The described study confirms our expectations that using multiple electrical parameters for detection of analyte with liquid gated FETs advances the accuracy of early diagnosis of HF. Nevertheless, the small size of GO flakes may increase variations in the aptasensor parameters. From that perspective, the use of functionalized monolayer graphene as the FET channel would be preferable. Moreover, a deeper insight in the signal transfer between the target, receptor, and graphene in such complex bionanohybrids as well as investigation of the effect of the complex matrix of body fluids is needed to turn the suggested approach into a versatile platform for real-time HF parameters diagnostics.

**Declaration of competing interest**

The authors declare that they have no known competing financial interests or personal relationships that could have appeared to influence the work reported in this paper.

**Data availability**

Data will be made available on request.

**Acknowledgments**

This research was supported in part by projects funded by the European Union's Horizon 2020 research and innovation programme: the IPANEMA, under grant agreement N° 872662 (https://doi.org/10.3030/872662) and ANTARES, under grant agreement No. 664387 (https://doi.org/10.3030/739570). Graphene chip development and characterization were supported by Russian Science Foundation under grant number 19-19-00401 (https://rscf.ru/en/project/19-19-00401/). Investigation of interactions of aptamer with cardiac markers and determination of analytical parameters were supported by Russian Science Foundation under grant number 19-73-10205 (https://rscf.ru/en/project/19-73-10205/). S.J. acknowledges the financial support of the Ministry of Science, Technological Development, and Innovations (grant N°. number 451-03-47/2023-01/200358). A.V.O. and P.I.N. were supported by the Ministry of Science and Higher Education of the Russian Federation, contract No. 075-15-2022-315. All the co-authors thank Dr. Dmitry Kireev (The University of Texas at Austin, USA) for providing array of rGO-FETs patterned on silicon wafer.